\documentstyle[prb,amsmath,aps,epsfig,twocolumn]{revtex}

\newcommand{\upd}{{\mathrm d}}
\newcommand{\eps}{\varepsilon}
\renewcommand{\phi}{\varphi}
\newcommand{\tf}{T_{\rm eff}}
\newcommand{\mtr}{m_{\rm tr}}
\newcommand{\tr}{t_{\rm rel}}

\begin{document}

\draft
\twocolumn[\hsize\textwidth\columnwidth\hsize\csname @twocolumnfalse\endcsname
\title{Nonequilibrium dynamics and fluctuation-dissipation relation
in a sheared fluid}

\author{Ludovic Berthier$^{1}$  and Jean-Louis Barrat$^2$ }

\address{$^1$CECAM, ENS-Lyon, 46, All\'ee d'Italie, 69007 Lyon, France}

\address{$^2$D\'epartement de Physique des Mat\'eriaux,
UCB Lyon 1 and CNRS, 69622 Villeurbanne, France}

\date{\today}

\maketitle

\begin{abstract}
The nonequilibrium dynamics of a binary Lennard-Jones mixture in a
simple shear flow is investigated by means of molecular dynamics
simulations. The range of temperature $T$ investigated covers both
the liquid, supercooled and glassy states, while the shear rate
$\gamma$ covers both the linear and nonlinear regimes of rheology.
The results can be interpreted in the context of a nonequilibrium,
schematic mode-coupling theory developed recently, which makes the
theory applicable to a wide range of soft glassy materials. The
behavior of the viscosity $\eta(T,\gamma)$ is first investigated.
In the nonlinear regime, strong shear-thinning is obtained, $\eta
\sim \gamma^{- \alpha(T)}$, with $\alpha(T) \simeq 2/3$ in the
supercooled regime. Scaling properties of the intermediate
scattering functions are studied. Standard `mode-coupling
properties' of factorization and time-superposition hold in this
nonequilibrium situation. The fluctuation-dissipation relation is
violated in the shear flow in a way very similar to that predicted
theoretically, allowing for the definition of an effective
temperature $\tf$ for the slow modes of the fluid. Temperature and
shear rate dependencies of $\tf$ are studied using density
fluctuations as an observable. The observable dependence of $\tf$
is also investigated. Many different observables are found to lead
to the same value of $\tf$, suggesting several experimental
procedures to access $\tf$. It is proposed that a tracer particle of
large mass $\mtr$ may play the role of an `effective thermometer'.
When the Einstein frequency of the tracers becomes smaller than the
inverse relaxation time of the fluid, a nonequilibrium equipartition
theorem holds with $\left\langle \mtr  v_z^2 \right\rangle =
k_{\rm B} \tf$, where $v_z$ is the velocity in the direction
transverse to the flow. This last result gives strong support to
the thermodynamic interpretation of $\tf$ and makes it
experimentally accessible in a very direct way.
\end{abstract}


\vspace*{-0.5cm}

\pacs{{\it L'immense thermom\`etre refl\'etait modestement
la temp\'erature du milieu ambiant}. \\
Vassili Axionov, {\it L'\^{\i}le de Crim\'ee}.}


\vskip2pc]

\narrowtext

\section{Introduction}
\label{intro}
 Glasses are usually defined by the fact that their
internal relaxation time is larger than the experimental time
scale. In simple molecular systems, the associated glass
transition temperature corresponds to  a very high viscosity,
making it difficult to investigate experimentally the rheological
properties of glassy systems. In complex fluids (e.g. colloids,
emulsions) it is however possible to reach a glassy situation,
in the sense of large relaxation times, with systems having
viscosities or shear moduli that allow for a rheological
investigation~\cite{larson}.
Such materials have been described as `soft glassy
materials' in the literature \cite{SGR}.

When it is quenched into its glassy state, a material is by
definition out of equilibrium. An important feature of this
nonequilibrium situation is the absence of time translation
invariance. Physical properties are a function of the time spent
in the glassy phase, or waiting time $t_w$. This is best seen
through the measurement of time-dependent correlations which
depend both on $t_w$ and on the time difference. These
dependencies on $t_w$ are usually described as aging
phenomena~\cite{struik}. They have been studied experimentally and
theoretically in great detail~\cite{aging_review}. Interestingly
for the present work, the aging behavior of several complex fluids
has recently been experimentally
investigated~\cite{luca,cloitre,laurence,bonn,lequeux,lequeux2,Clarke},
showing interesting similarities with other, more standard, glassy
systems.

By submitting the system to a homogeneous, steady shear flow, a
different kind of nonequilibrium situation is obtained. The steady
shear flow, characterized by the shear rate $\gamma$, creates a
{\em nonequilibrium steady state}, in which time translation
invariance is recovered~\cite{cukulepe}.
The shear flow can therefore be used as a
probe of the glassy system, with the convenient
feature of having  the shear rate
$\gamma$ rather than the waiting time $t_w$ as a control
parameter. In fact, the inverse shear rate introduces a time scale
in the problem, which plays a role similar to $t_w$.
Interesting phenomena are thus expected
to show up as soon as this new time scale competes with the relaxation
time of the fluid.
As a result, it is also interesting to study
the `supercooled' regime, which would correspond
to an equilibrated situation in the absence of a shear flow.
Moreover, this way of probing the nonequilibrium
properties of glassy systems is possibly more relevant
experimentally than the aging approach, at least in the
case of soft glassy materials.

Recently, a general scenario was proposed for glassy systems
subject to an external forcing~\cite{BBK}, based on the study of
simple `mean-field' models. The rationale of this approach is that
the equilibrium dynamics of these models is equivalent to the
`schematic' mode-coupling approach of slowing down in supercooled
liquids~\cite{schematic,kithwo,gotze}. The study of their
nonequilibrium dynamics can thus be seen as a {\it nonequilibrium}
schematic mode-coupling approach~\cite{bckm}. In this respect, it
is interesting to note that the mode-coupling theory was recently
extended to the aging regime~\cite{latz} with results which were
first predicted from the study of mean-field
models~\cite{leticia3,cuku}. To our knowledge, the mode-coupling
theory of supercooled fluid  has not yet been extended to fluids
under shear beyond the linear response of the supercooled regime,
i.e. at equilibrium. However, in analogy to what was done for
aging or supercooled systems, it seems sensible to bypass this
aspect and to carry out a direct comparison between the
predictions from mean-field theories and experimental or
simulation results, with the hope that the general scenario is
robust enough that the details of the system under study are
relatively unimportant.

Several earlier studies have been devoted to simulating fluid
under shear. Onuki and Yamamoto~\cite{Onuki} concentrated their
works on the  understanding of the dynamics at the molecular
level. Our study is focused on more global aspects, with the aim
of providing some experimentally testable predictions. Liu and
Nagel~\cite{Liunagel} also proposed to use the shear
rate as a relevant control parameter for jamming systems.
Liu and coworkers~\cite{Liulanger} investigated some aspects of the
nonequilibrium dynamics of a model similar to ours, with the
difference of being athermal (zero temperature) in the absence of shear.
Some of their results are closely related to
ours, as will be discussed in Sections~\ref{rheology} and \ref{FDT}.
More phenomenological approaches may be found in
Refs.~\cite{SGR,lequeux2,lequeux1}.

The aim of the present work is to check, on a realistic model
of the fluid, some of the predictions that emerged from
the earlier study of driven glassy systems~\cite{BBK}.
For completeness, the main results of this study will be
briefly recalled in section \ref{rappel}.
In section~\ref{model}, we describe our microscopic model
for the fluid under shear. Sections~\ref{rheology} and
\ref{correlations} describe our results for the dynamic
properties of the fluid, both at the macroscopic (rheological) and
at the microscopic (wavevector dependent correlations) scales.
Sections~\ref{FDT} and \ref{obs} are devoted to one of the most important
predictions of the mean-field scenario, namely the manner
in which the equilibrium fluctuation-dissipation
theorem is violated in such nonequilibrium systems and the resulting
notion of an effective temperature.
Section~\ref{propo} proposes a numerical realization
of a very simple experimental protocol to measure
the effective temperature through a nonequilibrium
generalization of the equipartition theorem.
We discuss experimental and theoretical consequences of our
results in Section~\ref{discussion}.

\section{Summary of the nonequilibrium mode-coupling results}
\label{rappel}

To make the paper self-contained, we briefly recall in this
Section the main results and predictions obtained within the
mean-field approach of Ref.~\cite{BBK}. In this paper, a simple
system, the $p$-spin mean-field model, was studied under
conditions that involved a constant energy input from an external
driving force, analogous to the effect of a shear flow on a fluid. The
resulting stationary nonequilibrium state was studied as a
function of the temperature $T$ and external drive,
the intensity of which will be,
for convenience, denoted by $\sigma$.

For $\sigma=0$, the model has an ideal glass transition
at a finite temperature $T_c$ where the relaxation time diverges.
For $T>T_c$, the
correlation functions are described by usual mode-coupling
integro-differential equations~\cite{kithwo},
and display the characteristic two-step
relaxation predicted by these equations,
the $\alpha$- and $\beta$-relaxations.
Below $T_c$, the system cannot equilibrate,
and displays aging behavior~\cite{cuku}.

Under a finite external drive, $\sigma \neq 0$, the system is
stationary at all temperatures. For $T \lesssim  T_c$, and in the
asymptotic limit $\sigma \to 0$, the  correlation functions retain
the characteristic two-step shape of the equilibrium system, with
an $\alpha$-relaxation time that depends on $\sigma$. For $T<T_c$,
this relaxation time diverges as $\sigma \to 0$, while for $T>T_c$
it goes to its equilibrium value in this limit.
Although the driving force strongly influences the $\alpha$-relaxation,
it does so by keeping the shape of the decay of the correlation
unchanged. This leads to the property that correlation functions
may be collapsed by a simple rescaling of the time.
This scaling property is analogous to the time-temperature
superposition property found at equilibrium, and we shall
denote this prediction by `time-shear superposition property'.

Based on a power dissipation argument, we were able for this
simple system to define quantities equivalent to the viscosity and
shear rate in a fluid under shear.  This viscosity was found to
have a characteristic shear-thinning behavior, decreasing for increasing
shear rates as a power law.
The shear-thinning
exponent is equal to $2/3$ for $T>T_c$, while for $T<T_c$, the
viscosity diverges as $\sigma \to 0$, and the shear-thinning
exponent appears to depend on temperature with
a value between 2/3 and 1.

The behavior of the effective temperature was also
investigated as a function of $T$ and $\sigma$. We recall here
that the effective temperature $\tf$
is {\it defined}, in a system invariant under time
translation,  by the relationship~\cite{leticia3}
\begin{equation}
 R(t) = - \frac{1}{k_B \tf(C)} \frac{\upd C(t)}{\upd t}.
 \label{definition}
\end{equation}
Here $R(t)$ and $C(t)$ are, respectively, a response function and
the associated correlation function. In equilibrium,  the
fluctuation-dissipation theorem can be written as $\tf(C)=T$. In the
sheared system, the effective temperature was found, for $T<T_c$, to be a
discontinuous function of $C$ in the limit $\sigma \to 0$. For
$C>q$, one has $\tf = T$ while for $C<q$, one finds
$\tf(C) = \tf > T$, denoting an effective temperature
larger than the bath temperature for long time scales~\cite{leticia3}.
The parameter $q$ is the plateau value of the correlation function:
it is called the Edwards-Anderson parameter in the literature
of disordered systems, or nonergodicity parameter
in the language of mode-coupling theory.
At finite shear, the effective temperature also has
an almost discontinuous behavior as a function of $C$. For
$T>T_c$, the discontinuity decreases and vanishes when $\sigma \to
0$, so that equilibrium behaviour is recovered in this limit.

An important outcome of theoretical studies
of the concept of effective temperature for the slow modes
of a glassy system, is that the value of the effective temperature
is independent, at the mean-field level, of which observable
is chosen to compute correlation and response
functions~\cite{latz,leticia3,doudou,leticia4}.
This prediction was recently challenged within a
simple trap model~\cite{suzanne}, with negative results.
We will see in Section~\ref{FDT} that this important
prediction is nicely verified in our realistic model.

The simple mean-field model studied in Ref.~\cite{BBK} does
not allow any discussion of spatial dependency.
However, characteristic features of the spatial
dependence usually expected from the solution of mode-coupling
like equations will be investigated in Section~\ref{spatial}.

\section{Model and details of the simulation}
\label{model}

The system simulated in this work is a 80:20 mixture
of $N=2916$ Lennard-Jones particles of types $A$ and $B$, with interaction
\begin{equation}
V(\boldsymbol{r_{\alpha \beta}}) = 4 \eps_{\alpha \beta} \left[
\left(\frac{\sigma_{\alpha \beta}} {|\boldsymbol{r_{\alpha \beta}}|}
\right)^{12} -
\left(\frac{\sigma_{\alpha \beta}}{|\boldsymbol{r_{\alpha \beta}}|}
\right)^{6}
\right],
\end{equation}
where $\alpha$ and $\beta$ refer to the two species $A$ and $B$.
Interaction parameters are chosen to prevent crystallization~\cite{kob}.
In all the paper, the length, energy and time units are the standard
Lennard-Jones units $\sigma_{AA}$ (particle diameter),
$\eps_{AA}$ (interaction energy), and $\tau_0 =
(m_A\sigma_{AA}^2/ \eps_{AA})^{1/2}\,$ \cite{note1}, where $m_A$
is the particle mass and the subscript $A$ refers to the majority
species. Particles have equal masses, and the
interaction parameters are $\eps_{AB} = 1.5 \, \eps_{AA}$,
$\eps_{BB} = 0.5 \, \eps_{AA}$, $\sigma_{BB}= 0.88
\, \sigma_{AA}$, $\sigma_{AB} = 0.8 \,  \sigma_{AA}$. With these
interaction parameters between species, equilibrium
properties of the system have been fully
characterized~\cite{kob}.
At the reduced density $\rho=1.2$, where all our
simulations are carried out, a `computer glass transition' is
found in the vicinity of $T_c \sim 0.435$. The slowing down of the
dynamics, $T \gtrsim 0.45$, is correctly described by mode-coupling
theory, which breaks down when lowering further the temperature,
$T \lesssim 0.45$~\cite{kob}.
The aging behavior of the system below this
temperature has also been characterized extensively, including the
violation of the FDT in the glassy phase~\cite{barratkob}.

In order to study the system under a steady shear flow, the classical
Newton equations are replaced by the so- called
SLLOD equations~\cite{evans}
\begin{equation}
\begin{aligned}
\frac{\upd \boldsymbol{r_i}}{\upd t} = & \frac{\boldsymbol{p_i}}{m_i}
+ \boldsymbol{r_i}\cdot \boldsymbol{\nabla v}, \\
\frac{\upd \boldsymbol{p_i}}{\upd t} = & - \sum_{j \neq i}
\frac{ \partial V(\boldsymbol{r_{ij}})}{\partial \boldsymbol{r_{ij}}}
- \boldsymbol{p_i}
\cdot \boldsymbol{\nabla v},
\label{sllod}
\end{aligned}
\end{equation}
where  $(\boldsymbol{p_i},\boldsymbol{r_i})$ are the momentum and
position of particle $i$, respectively. These equations are
integrated by a standard leapfrog algorithm~\cite{AT87}, where
time is discretized. The value $\Delta t = 0.01$ was used
throughout the simulations. Lees-Edwards boundary
conditions~\cite{AT87} are used in a cubic simulation box of
linear size $L=13.4442$. With these boundary conditions the flow
is homogeneous and no instability, such as e.g. shear-banding, was
observed in our simulations~\cite{note0,roro}. Note also that this is
the shear rate $\gamma$ which is controlled in our simulations.
The velocity gradient is in the $y$ direction, and the fluid
velocity in the $x$ direction, i.e. $\boldsymbol{v} = \gamma y
\,\, \boldsymbol{e_{x}}$. Constant temperature conditions are
ensured by a simple velocity rescaling of the $z$ component of the
velocities, at each time step \cite{note2}.

The shear rate ${\gamma}$ naturally introduces a new time scale
$\gamma^{-1}$ into the problem. Obviously, a simulation involving
a steady shear state is possible only if the available simulation
time is significantly larger than $\gamma^{-1}$. This limits our
study to shear rates larger than typically $10^{-4} \tau_0$,
corresponding to $10^6$ time steps. The temperature range studied
here is $T \in [0.15,0.6]$, which covers the liquid and glassy
phases, while the shear rates are $\gamma \in [10^{-4},0.1]$ which
covers the linear and nonlinear response regimes of rheology. All the data
presented in the following, even those below $T_c$, are obtained
in a stationary state. This is unusual for systems below the glass
transition temperature, which normally exhibit a non-stationary
aging behavior.

\section{Macroscopic behavior: Nonlinear rheology}
\label{rheology}

In this Section, we study the rheological behavior
of the system. Our results are
very similar to that obtained in many different soft condensed
matter systems, making the Lennard-Jones mixture ---which was
originally designed as a model for simple metallic glasses--- a
reasonable one for more complex systems.

Before discussing our results, let us briefly recall some possible
behaviors for the flow curves (stress $\sigma$ as a function of
shear rate) obtained in complex fluids. These results are often
represented in the phenomenological form~\cite{larson,SGR}
\begin{equation}
\sigma \simeq  \sigma_0 + a \gamma^{n}.
\label{phenom}
\end{equation}
The case $n=1$ corresponds to Bingham fluids, with a yield stress
$\sigma_0$, which vanishes for Newtonian systems. For $n<1$,
Eq.~(\ref{phenom}) is often called the Hershel-Bulkeley law~\cite{larson}.
For zero yield stress ($\sigma_0=0$), Eq.~(\ref{phenom})
describes a `power-law fluid', where the viscosity $\eta$ decreases
with increasing the shear rate as a power law,
$\eta \sim \gamma^{-\alpha}$, $\alpha \equiv 1-n$.
Although these type of
flow curves are found in many different complex systems, they are
not usually supported by a theoretical basis, and should therefore
be interpreted as a convenient representation of the results on a
limited range of shear rates.
In our system, the microscopic stress  $\sigma_{xy}$  is defined
by the usual formula
\begin{equation}
\sigma_{xy} \equiv \frac{1}{L^3} \left( \sum_{i=1}^{N} -
\frac{p_{ix} p_{iy}}{m_i} + \sum_{i=1}^{N} \sum_{j >  i}
r_{ijx} \frac{\partial V(\boldsymbol{r_{ij}})}{\partial r_{ijy}}
\right). \label{stressmicro}
\end{equation}
The shear rate dependent viscosity is defined by the ratio
\begin{equation}
\eta \equiv \frac{\sigma_{xy}}{\gamma}.
\end{equation}

\begin{figure}
\begin{center}
\psfig{file=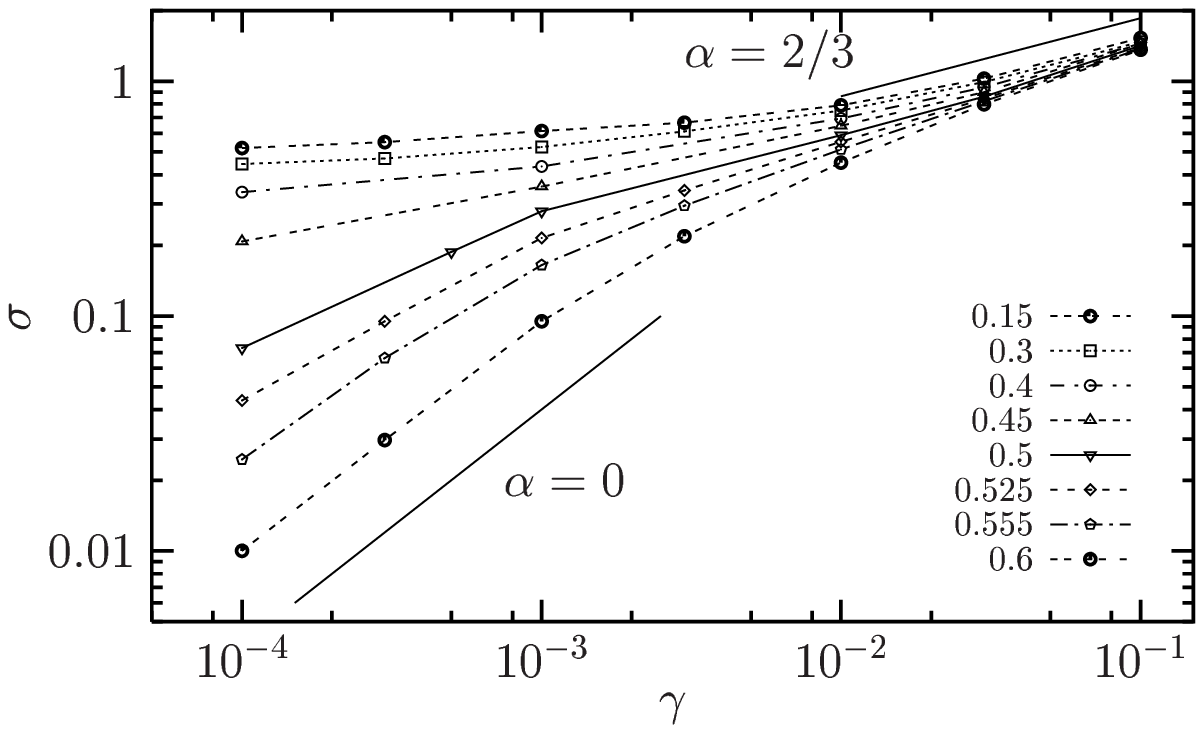,width=8.5cm,height=6cm}
\psfig{file=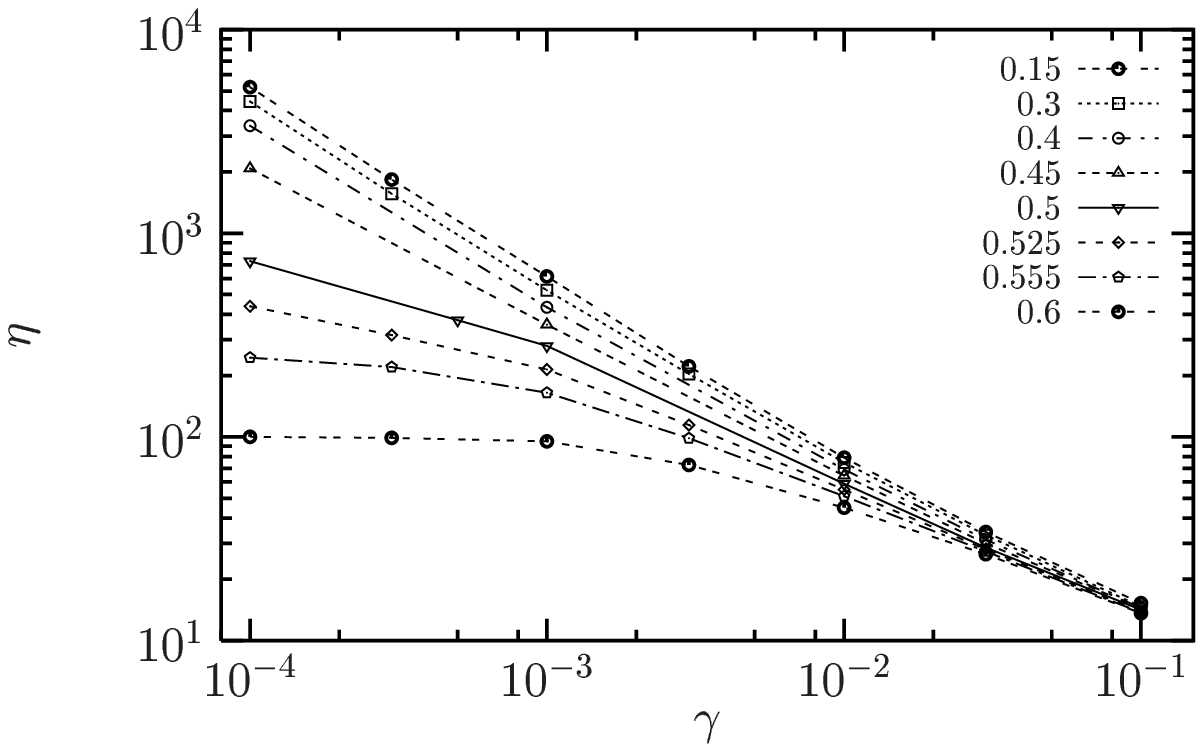,width=8.5cm,height=6cm}
\caption{Top: Flow curves $\sigma(\gamma)$ for various temperatures.
Full lines are Newtonian behavior ($\alpha=0$) and shear-thinning
behavior with $\alpha=2/3$.
Bottom: Flow curves presented in the alternative
form $\eta(\gamma)$, for the same temperatures.}
\label{flow2}
\end{center}
\end{figure}

\begin{figure}
\begin{center}
\psfig{file=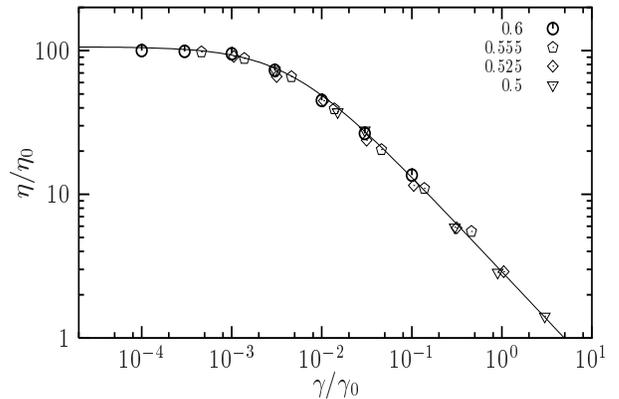,width=8.5cm,height=6cm}
\caption{Scaling
behavior of the flow curves $\eta(\gamma)$ for various
temperatures above $T_c$. The full line is a fit to
equation~(\ref{fit}).} \label{viscoresc}
\end{center}
\end{figure}

Figure~\ref{flow2} presents the flow curves $\sigma(\gamma,T)$
for various temperatures. The same data are also represented in the
alternative form of viscosity versus shear rate. Two
different regimes can be distinguished in these figures. For
$T>T_c$, the behavior becomes Newtonian at low shear rates,
\begin{equation}
\sigma \simeq \eta_0(T) \gamma,
\end{equation}
 where
\begin{equation}
\eta_0(T)= \lim_{\gamma \to 0} \eta(\gamma,T>T_c)
\end{equation}
is the viscosity of the fluid in the linear regime.
We have checked that it is proportional to the equilibrium
relaxation time of the fluid~\cite{Onuki},
and it is found to diverge as the temperature is lowered
as a power law $\eta_0(T) \sim (T-T_c)^{-2.45}$, in good agreement
with equilibrium simulations~\cite{kob}.
Figure~\ref{viscoresc} shows that
in this temperature regime, all the data  can be rescaled onto a
master curve of the form
\begin{equation}
\eta(\gamma,T) = \frac{\eta_0(T)}{ \big[ 1+\gamma/\gamma_0(T)
\big]^{2/3}}.
\label{fit}
\end{equation}
$\gamma_0(T)$ is not a free parameter, but has the form
$\gamma_0(T) \sim (T - T_c)^{-2.45\times 3/2}$, which ensures that
the divergence of the viscosity at $T_c$ is a  power law
$\eta(\gamma,T_c) \sim \gamma^{-2/3}$.

Equation~(\ref{fit}) describes a crossover between a Newtonian
behavior at small shear rates and a power law behavior with $\eta
\sim \gamma^{-2/3}$ at higher shear rates. Remarkably, the same
behavior, with the same exponent $2/3$ were predicted by the
mean-field theory of Ref.~\cite{BBK}. This coincidence, however,
is not necessarily significant. A similar shear-thinning behavior,
$\eta \sim \gamma^{-\alpha}$, with exponents in the range $\alpha
=0.5-1.0$ is obtained in many different soft systems, such as
suspensions and concentrated polymer
solutions~\cite{larson,ferry}. Confined polymer layers under
constant normal load~\cite{robbins} were also found to exhibit
very similar shear-thinning behavior. In our case, the data could
probably be fitted with a slightly different value for the
exponent, and the choice of $2/3$ is motivated by theoretical
results~\cite{BBK}.

\begin{figure}
\begin{center}
\psfig{file=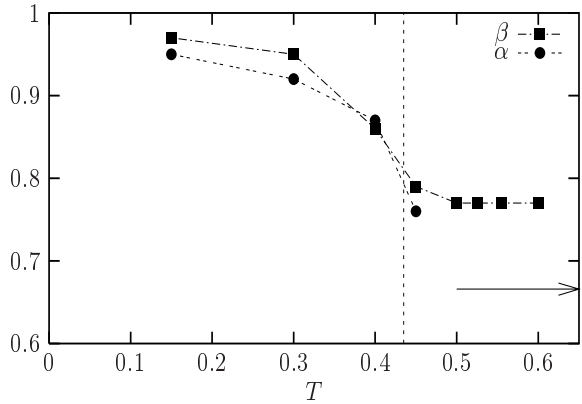,width=8.5cm,height=6cm}
\caption{Evolution
of the effective shear-thinning exponent $\alpha(T)$ and of the
stretching exponent $\beta(T)$. The horizontal arrow shows the value
$\alpha=2/3$ obtained above $T_c$. The vertical
dashed line is at $T_c=0.435$.}
\label{beta}
\end{center}
\end{figure}

\begin{figure}
\begin{center}
\psfig{file=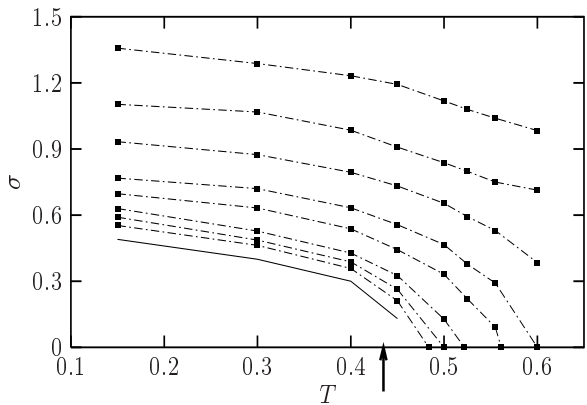,width=8.5cm,height=6cm}
\caption{The $(\sigma,T)$ plane of the jamming phase diagram.
The dashed curves are the viscosity contour plots.
$\eta=20$, 30, 50, 100, 200, 500, 1000 and 2000 (from top
to bottom). The full line represents the
value of the yield stress $\sigma_0(T)$, extrapolated from
Eq.~(\ref{phenom}). Arrow marks the mode-coupling temperature $T_c$.}
\label{jamming}
\end{center}
\end{figure}

At low temperatures, the curves shown in Fig.~\ref{flow2}
present a crosssover between two shear-thinning regimes.
For relatively high shear rates, they can be described
by the same shear-thinning exponent as above $T_c$, $\alpha=2/3$,
but at low shear rate, the shear-thinning behavior
is more marked, $\alpha > 1$.
One can define an effective shear-thinning exponent
from the slope of the curves in Fig.~\ref{flow2}
($\alpha= \upd \ln \eta / \upd \ln \gamma)$).
It crosses over from $\alpha \sim 2/3$ at relatively high shear rate,
to a larger value for the lowest shear rate investigated here.
The temperature dependence of the latter
is reported in Fig.~\ref{beta},
which shows that $\alpha(T)$ saturates to 1 when $T\to 0$,
which seems to indicate that the system has a finite
yield stress in this limit only.

Although our results are obtained on 3 decades of shear rates, we
cannot report a definite functional form for the flow curves.
An alternative view of the low temperature flow curves is that
the effective shear-thinning exponent will saturate to the value
$\alpha(T) = 1$ in the (numerically unreachable) very
low shear rate limit, indicating the existence of
a yield stress $\sigma_0(T) \equiv \lim_{\gamma
\to 0} \sigma(\gamma,T)$ for the system at finite temperatures.
Hence, we also tried to use Eq.~(\ref{phenom}), in order to extract a
possible value of $\sigma_0(T)$. Note that this is made
possible by the fact that Eq.~(\ref{phenom}) has more free
parameters than a single power law. Fits are thus satisfactory,
and the values of the yield stress obtained at different
temperatures in this manner are reported in Fig.~\ref{jamming}.
This `jamming phase diagram' is very similar to the one proposed
by Liu and Nagel~\cite{Liunagel}, and studied experimentally in
Ref.~\cite{weitz}. It is also an interesting way of thinking of
the glass transition, and as such was theoretically investigated
in Ref.~\cite{BBK}. The `jammed phase' would correspond here to
the low-$\sigma$, low-$T$ part of the diagram.

\section{Microscopic correlation functions}
\label{correlations}

The shear rate may be viewed as a new control
parameter to access the glassy phase. As discussed earlier, one
may expect scaling properties of the correlation functions as a
function of shear rate to be similar to those observed when
temperature is decreased. This basic remark was already
made in Ref.~\cite{Onuki}, but we now have
a clear theoretical context in which these results
may be understood~\cite{BBK}.
In this Section, we briefly consider the
influence of the shear rate on static correlations, and  investigate
next the scaling properties of the time-dependent correlation
functions.

\subsection{Static correlation functions}

Static correlations may be characterized through the structure
factors of the fluid, defined as
\begin{equation}
S(\boldsymbol{k}) = \frac{1}{N} \sum_{j=1}^N \sum_{l=1}^{N}
\left\langle
\exp  \bigg(  i \boldsymbol{k} \cdot \big[ \boldsymbol{r_j}(t_0)-
\boldsymbol{r_l}(t_0) \big] \bigg)  \right\rangle.
\end{equation}

\begin{figure}
\begin{center}
\psfig{file=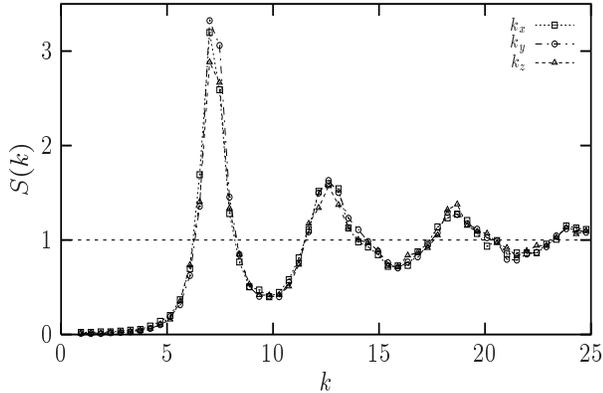,width=8.5cm,height=6cm}
\caption{Structure factor at fixed temperature, $T=0.3$, and
shear rate, $\gamma=10^{-3}$, for the different orientations with
respect to the flow.}
\label{struc2}
\end{center}
\end{figure}

We show first in Fig.~\ref{struc2} the structure
factor for $\gamma=10^{-3}$, and $T=0.3$
in the three spatial directions.
It can be seen that the flow, somehow surprisingly,
introduces no obvious anisotropy
in this structural quantity~\cite{Onuki}.
This can be explained by the fact that we are
working at shear rates which are small, $\gamma \tau_0 \ll 1$.
A structural anisotropy would probably be present for larger shear
rates, $\gamma \tau_0 \gtrsim 1$.

\begin{figure}
\begin{center}
\psfig{file=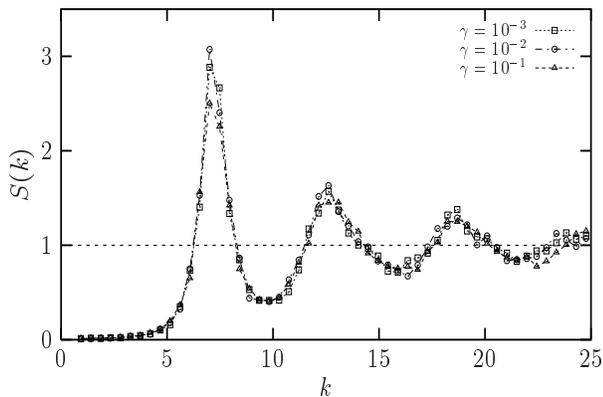,width=8.5cm,height=6cm}
\caption{Structure factor at fixed temperature $T=0.3$ and different shear
rates. The relaxation time increases by $\sim 2$ orders of magnitude
on the same shear rate interval.}
\label{struc}
\end{center}
\end{figure}

We present next in Fig.~\ref{struc} the evolution of $S(k_z)$ for
different shear rates at constant temperature $T < T_c$. The
`glassy' state is thus approached when the shear rate is lowered.
We shall see below that for these values of the shear rate, the
relaxation time increases by nearly two orders of magnitude. Only
weak changes in the static structure factor are observed in the
same shear rate window. The situation is thus strongly reminiscent
of the observation made at equilibrium that no structural sign of
the glass transition can be observed~\cite{kob,reviewglass}.

\subsection{$\boldsymbol \alpha$-relaxation and `time-shear superposition'}

To investigate the dynamics of the fluid under shear, we focus on
the number density fluctuations
$\langle \delta \rho (\boldsymbol{k},t) \delta \rho (-\boldsymbol{k},t')
\rangle $
where
\begin{equation}
\delta \rho (\boldsymbol{x},t) = \left\langle
\frac{1}{N} \sum_{j=1}^N \delta \big( \boldsymbol{x} -
\boldsymbol{r_j} (t) \big) - \rho \right\rangle
\end{equation}
is the density fluctuation at point $\boldsymbol{x}$ and time
$t$, $\rho$ being
the average density of the fluid. For simplicity, we will limit
ourselves to ${\boldsymbol k}$ vectors orthogonal to the flow
directions, for which the overall motion of the fluid does not
affect the correlation function directly.

\begin{figure}
\begin{center}
\psfig{file=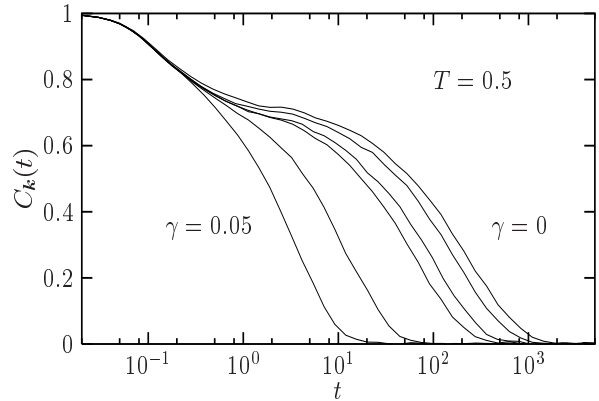,width=8.5cm,height=6cm} \\
\psfig{file=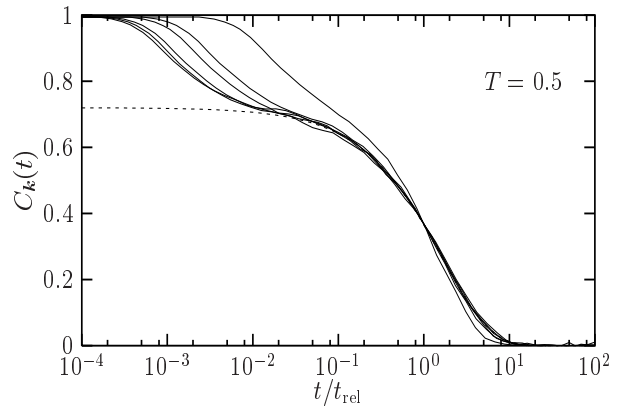,width=8.5cm,height=6cm}
\caption{Top:
Correlation functions for $T=0.5>T_c$ and different values of
the shear rate, $\gamma = 0$, $10^{-4}$,
$5 \cdot 10^{-4}$, $10^{-3}$, $10^{-2}$, $5 \cdot 10^{-2}$
(from right to left).
Bottom: The slow decay of the correlation function
can be collapsed if the time is rescaled by $t_{\rm rel}(\gamma)$.
The dashed line is a fit to a stretched exponential form, with
an  exponent $\beta = 0.77$.}
\label{corr}
\end{center}
\end{figure}

\begin{figure}
\begin{center}
\psfig{file=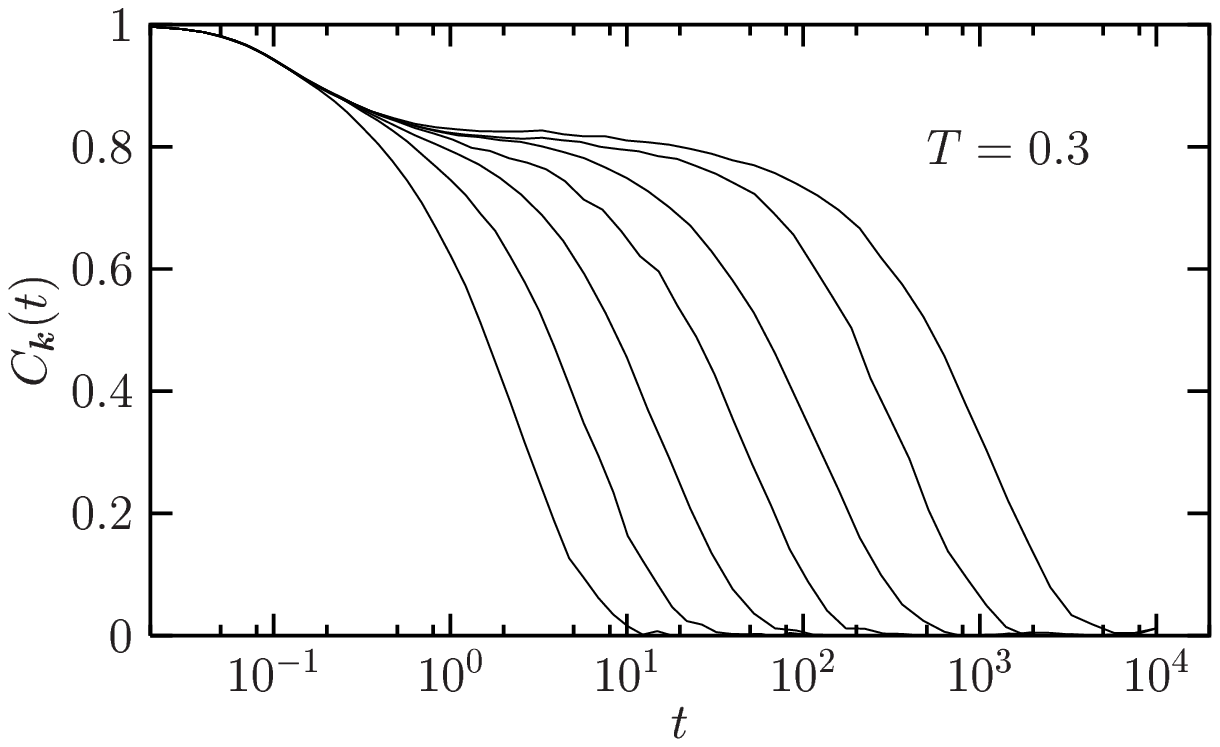,width=8.5cm,height=6cm} \\
\psfig{file=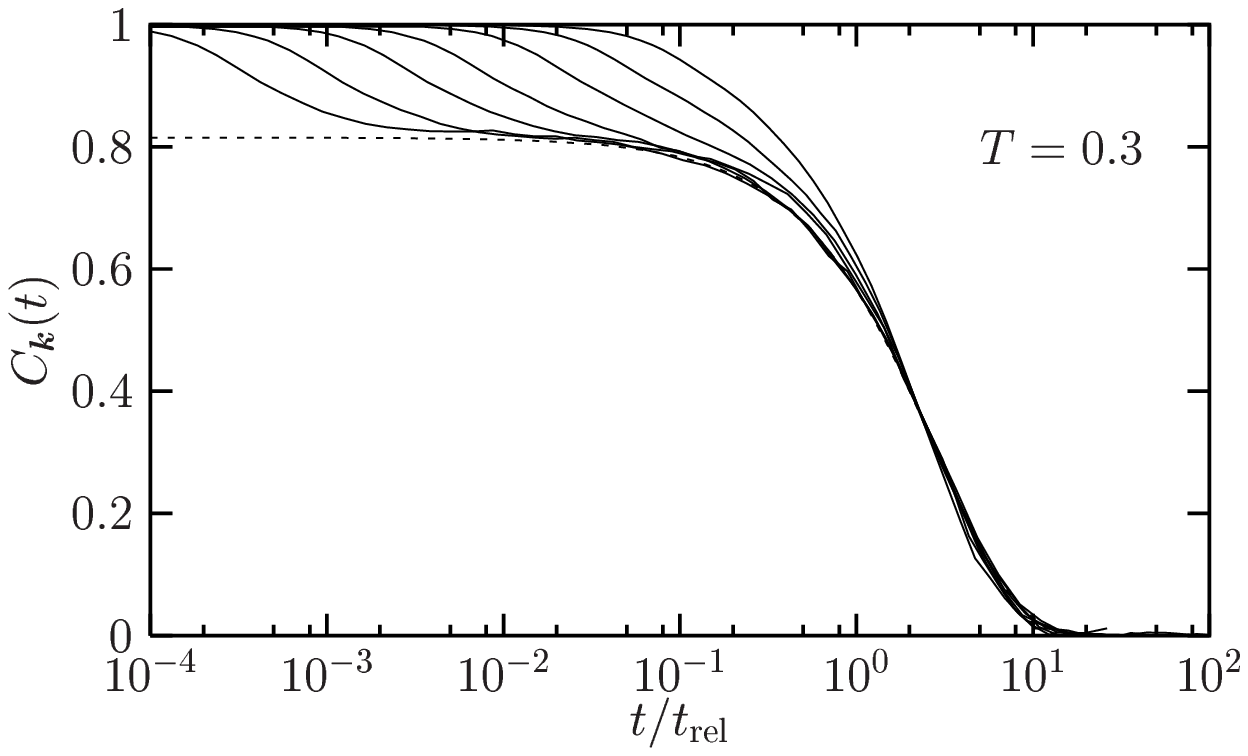,width=8.5cm,height=6cm}
\caption{Top:
Correlation functions for $T=0.3<T_c$ and different values of
the shear rate, $\gamma=10^{-4}$, $3 \cdot 10^{-4}$, $10^{-3}$,
$3 \cdot 10^{-3}$, $10^{-2}$, $3 \cdot 10^{-2}$,
$10^{-1}$ (from right to left).
Bottom: The slow decay of the correlation function
can be collapsed if the time is rescaled by $t_{\rm rel}(\gamma)$.
The dashed line is a fit to a stretched exponential form, with
an  exponent $0.95$.}
\label{corr3}
\end{center}
\end{figure}

To discuss the dynamic behavior of the system,
we concentrate on the self-part of the intermediate scattering function,
defined by
\begin{equation}
C_{\boldsymbol{k}}(t) = \frac{1}{N_A} \sum_{j=1}^{N_A} \left\langle
\exp  \bigg(  i \boldsymbol{k} \cdot \big[ \boldsymbol{r_j}(t+t_0)-
\boldsymbol{r_j}(t_0) \big] \bigg)  \right\rangle.
\label{LEdef}
\end{equation}
We have also computed this function for the minority species, $B$,
with very similar results, that will not be discussed here.
These correlation functions are displayed for two different temperatures
and several shear rates in Figs.~\ref{corr} and \ref{corr3}.
The wavector is $\boldsymbol{k} = 7.47 \boldsymbol{e_z}$, which corresponds
to the peak of the structure factor, see Fig.~\ref{struc2}.
The overall shape is very similar to what is usually found in supercooled
systems, namely a first `microscopic' relaxation which is
independent of the control parameters, followed by a slow approach
to a `plateau' which we will describe as `$\beta$-relaxation' in
analogy with the mode-coupling terminology. The decay beyond the
plateau, or terminal relaxation, will be described as
`$\alpha$-relaxation'. Obviously, the $\alpha$-relaxation time
is a strongly decreasing function of shear rate~\cite{heyes}.

It is a theoretical prediction that the slow decay of the
correlation function may be collapsed by a simple rescaling of the
time, the `time-shear superposition property'~\cite{BBK}.
As in equilibrium~\cite{kob}, the relaxation
time is defined from the relation
$C_{\boldsymbol{k}}(\tr) \equiv e^{-1}$.
The scaling form
\begin{equation}
C_{\boldsymbol{k}}(t) =  F_{\boldsymbol{k}}
\left( \frac{t} {\tr(\gamma)} \right),
\label{superposition}
\end{equation}
where the function $F_{\rm \boldsymbol{k}}(x)$
is a master function that may depend on
temperature, is tested in Figs.~\ref{corr} and \ref{corr3}.
It can be seen that such a
scaling is indeed nicely obeyed for the $\alpha$-relaxation
of the correlation functions. The scaling function is
correctly described by a stretched exponential,
\begin{equation}
F_{\boldsymbol{k}}(x) \simeq  \exp \left( -  x^{\beta(T)} \right),
\label{stretched}
\end{equation}
with the stretching exponent $0 < \beta(T) < 1$.
The temperature evolution of $\beta(T)$ is shown in Fig.~\ref{beta},
which demonstrates that the decay becomes less stretched when the temperature
is lowered. The stretching exponent seems to saturate
to the value $1$ as $T$ decreases to $0$.
In this figure,  a close relationship between this stretching of
the correlation function and the shear-thinning exponent is also
clearly apparent. This is physically plausible:  $\beta=1$ corresponds to
a pure exponential form, i.e. a single relaxation time;  $\beta <
1$, on the other hand, is associated with  a  broader distribution of
relaxation times. If $\beta = 1 = \alpha$,
there is  a single relevant time
scale, and the scaling $\eta \sim \gamma^{-1} \sim \tr$
shows that this time
scale is simply given by the inverse of the shear rate, as
would be obtained by a naive dimensional analysis.
This suggests, as an interesting corollary, that a shear-thinning
exponent $\alpha < 1$ implies the existence of a broad distribution of
time scales.

\subsection{Spatial dependence and $\boldsymbol \beta$-relaxation}
\label{spatial}

\begin{figure}
\begin{center}
\psfig{file=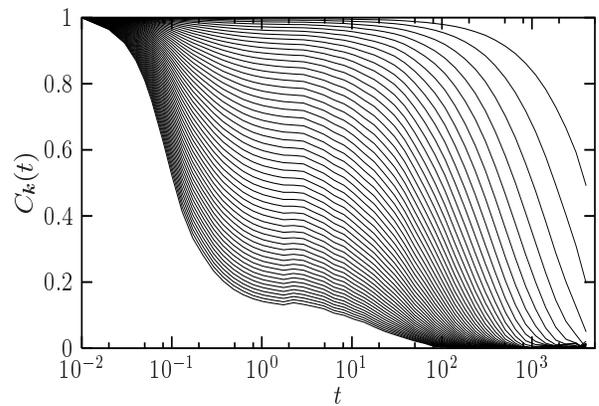,width=8.5cm,height=6cm}
\caption{Correlation
function for $T=0.3$, $\gamma=10^{-3}$, and different values of the
wavevector regularly spaced between $k=0.93$ and $k=24.76$
(from top to bottom).} \label{corrk}
\end{center}
\end{figure}

The wavevector dependence of the number-number correlation
function is displayed in Fig.~\ref{corrk}. As usual~\cite{kob},
the overall shape is qualitatively the same for all wavevectors,
the plateau value and the stretching exponent
being simple functions of the wavevector.

In mode-coupling theories~\cite{gotze,latz},
an interesting and easily testable
property of the correlation function is the so-called
`factorization property' which characterizes the spatial
dependence in the $\beta$-relaxation region. This factorisation
property can be written, for the correlation function under study,
in the form $C_{\boldsymbol{k}}(t) = q_k + h_k f(t)$,
where $q_k$ is the wavevector-dependent Edwards-Anderson
parameter, and $f(t)$ a universal function, i.e.
independent of the chosen correlation function.
More generally, it implies that in the $\beta$-regime the ratio
\begin{equation}
R_\phi(t) = \frac{\phi(t)-\phi(t'')}{\phi(t')-\phi(t'')},
\label{fact}
\end{equation}
where $\phi(t)$ is any slow observable [e.g.
$C_{\boldsymbol{k}}(t)$], and $t,t'$ and $t''$ are in the
$\beta$-relaxation window, is independent of the
observable~\cite{kob,barratkob}.

\begin{figure}
\begin{center}
\psfig{file=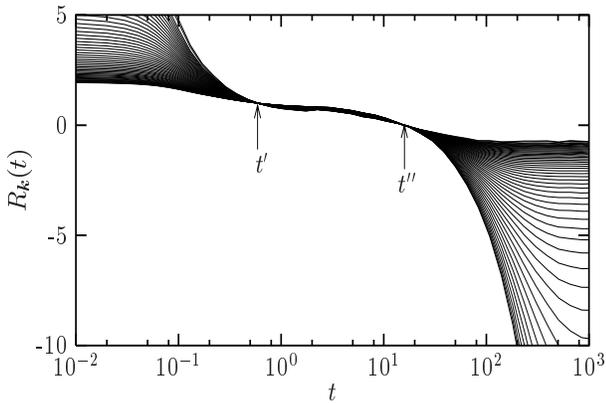,width=8.5cm,height=6cm}
\caption{The correlation functions of Fig.~\ref{corrk}
are rescaled in the $\beta$-regime using the
mode-coupling-called `factorization property', Eq.~(\ref{fact}).
Times are $t'=0.57$ and $t''=14$, as indicated
by the arrows.}
\label{corrkfact}
\end{center}
\end{figure}

This prediction is tested in Fig.~\ref{corrkfact}, which shows
that the ratios (\ref{fact}) obtained for different wavevectors are
indeed independent of the wavevector.  Again, these data are very
similar to what is observed at equilibrium~\cite{kob}, and in the aging
regime~\cite{barratkob}.

\section{Effective temperatures}
\label{FDT}

\subsection{Existence of an effective temperature}

Nonequilibrium mean-field or mode-coupling
theories have shown that whereas correlation functions
were the only relevant quantities to be studied at equilibrium,
it is of primary interest to separatly study
susceptibilities in out of equilibrium
situations~\cite{BBK,latz,leticia3,cuku}.
At equilibrium, these two families of dynamic functions
are related by fluctuation-dissipation theorems (FDT).
In nonequilibrium situations, FDT is a priori not satisfied.
However, it now is clear that the study of the corrections
to the FDT are relevant~\cite{aging_review}.
Recent theories of nonequilibrium systems have indeed introduced the
notion of effective temperature associated with the modification
of the fluctuation-dissipation theorem~\cite{leticia3,leticia4}.
This effective temperature is defined by the ratio of the response
to correlation function in the following way. Consider two
physical observables $O(t)$ and $O'(t)$. The correlation function
$C_{OO'}(t)$ is defined as
\begin{equation}
C_{OO'}(t) = \langle O(t+t_0) O'(t_0) \rangle -
\langle O(t_0) \rangle \langle O'(t_0) \rangle,
\end{equation}
where $t_0$ is an arbitrary initial time and
$\langle \cdots \rangle$ indicates an average
over different initial times $t_0$.
The conjugated response function is defined by
\begin{equation}
R_{OO'}(t) = \frac{\delta \langle O(t+t_0)
\rangle}{\delta h_{O'}(t_0)},
\end{equation}
where $h_{O'}$ is the field thermodynamically conjugated
to $O'(t)$.
In a system at thermodynamic equilibrium, the FDT reads
\begin{equation}
R_{OO'}(t) = -\frac{1}{T} \frac{\upd C_{OO'}(t)}{\upd t}
\label{FDTeq}
\end{equation}
The more easily accessible physical quantity is the susceptibility
\begin{equation}
\chi_{OO'}(t) = \int_0^t \upd t' \, R_{OO'}(t')
\end{equation}
which is obtained by applying a small, constant field $h_{O'}$ in the
time interval  $[ 0, t ]$. In the linear response regime, $h_{O'}
\to 0$, one gets
\begin{equation}
\chi_{OO'}(t) \simeq \frac{\langle O(t) - O(0) \rangle}{h_{O'}}
\label{repons}
\end{equation}
The equilibrium FDT thus implies a linear relation
between the susceptibility and the correlation, namely
\begin{equation}
\chi_{OO'}(t) = \frac{1}{T} \big( C_{OO'}(0) -C_{OO'} (t)
\big).
\end{equation}

In a system out of equilibrium, the FDT is not expected to hold.
However, in a system invariant under time translation, a modified
version of Eq.~(\ref{FDTeq}) can be used to {\it define}
the functions $\tf^{OO'}(C_{OO'})$ through~\cite{leticia3}
\begin{equation}
R_{OO'}(t) = -\frac{1}{\tf^{OO'}(C_{OO'})} \frac{\upd
C_{OO'}(t)}{\upd t} \label{FDTratio}.
\end{equation}
The $\tf^{OO'}(x)$ are a priori arbitrary functions of their
argument, and may depend on which observables $O(t)$ and $O'(t)$
are under study.
However, {\it they
can be measured} by following the same linear
response procedure as in the equilibrium case, so that
\begin{equation}
\begin{aligned}
\chi_{OO'}(t) = & \int_0^t \upd t' \left(
 -\frac{1}{\tf^{OO'}(C_{OO'})} \frac{\upd
C_{OO'}(t')}{\upd t'} \right)\\
= & \int_{C_{OO'}(t)}^{C_{OO'}(0)}  \frac{\upd x}{\tf^{OO'} (x)}
\end{aligned}
\label{FDTneq}
\end{equation}
The existence of an effective temperature is thus {\it
demonstrated} if a straight line is obtained in a
susceptibility-correlation plot parameterized by the time. As was
shown in several cases~\cite{leticia4}, the same system may
exhibit different temperatures on different time scales. If this
is the case, the parametric plot will consist in several different
straight lines, the slope of each line determining an effective
temperature. Cases are also known \cite{Quantum,Granular} in which
the parametric plot is ill-defined at short times (e.g. because of
an oscillatory behavior of the correlation function) but yields a
well-defined effective temperature on longer time scales. It is
therefore generally {\it incorrect} to define an effective temperature
through the approximate expression $\chi_{OO'}(t\to
\infty)/C_{OO'}(0)$, as is sometimes done~\cite{Liulanger}. This
corresponds to taking the average slope of the parametric plot as
an effective temperature, therefore losing all information on the
time scale dependence of the effective temperature.

Obviously, the introduction of an effective temperature is of
`thermodynamic' interest only if this quantity
is actually independent of the
observables $O(t)$ and $O(t')$ under consideration, $\tf^{OO'}
\equiv \tf$.
This is true at the mean-field level~\cite{latz,leticia3,doudou},
and this question will be investigated in detail below.

As a first investigation of the existence and behavior
of the effective temperature and fluctuation-dissipation
relation, we consider the case of single particle
density fluctuations, for which the corresponding
correlation function has been
studied in detail above. Corresponding
observables are in this case
\begin{equation}
O(t) = \frac{1}{N_A} \sum_{j=1}^{N_A} \eps_j
\exp \big( i \boldsymbol{k} \cdot \boldsymbol{r_j}(t) \big),
\label{obs1}
\end{equation}
and
\begin{equation}
O'(t) = 2 \sum_{j=1}^{N_A} \eps_j
\cos \big( \boldsymbol{k} \cdot \boldsymbol{r_j}(t) \big),
\label{obs2}
\end{equation}
where $\eps_j = \pm 1$ is a bimodal random variable
of mean 0. The relation
\begin{equation}
C_{\boldsymbol{k}}(t) = \overline{C_{\rm OO'}(t)},
\end{equation}
where the horizontal line means an average over the realizations of
$\{\eps_j\}$, easily follows.
To compute the susceptibility, a force
\begin{equation}
\boldsymbol{F_j}(\boldsymbol{k},t) = -
\frac{\partial}{\partial \boldsymbol{r_j}}
\bigg( - h_{O'} O' (t) \bigg)
\end{equation}
is exerted on each $A$-particle.
The response function $\chi_{\boldsymbol{k}}(t)$ [see Eq.~(\ref{repons})]
is computed by averaging the response obtained in different realizations
of the $\{\eps_j\}$.

\begin{figure}
\begin{center}
\psfig{file=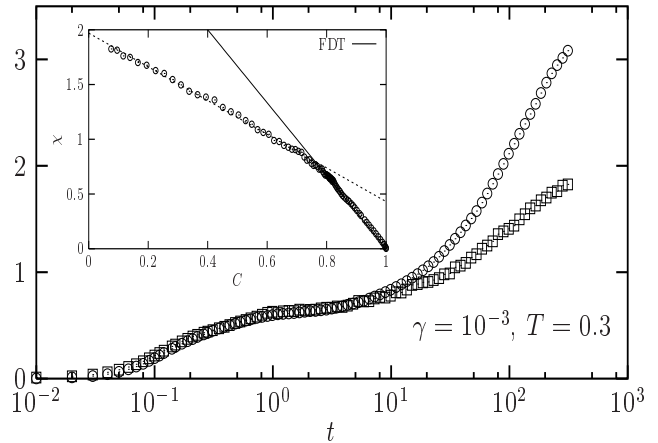,width=8.5cm,height=6cm}
\caption{Susceptibility $\chi_{\boldsymbol{k}}(t)$ (squares)
and $[1-C_{\boldsymbol{k}}](t)]/T$ (circles) versus time, for
$\boldsymbol{k} = 7.47 \boldsymbol{e_z}$,
$T=0.3$, and $\gamma=10^{-3}$.
The two curves are superposed when the equilibrium FDT is satisfied.
Inset: Parametric susceptibility versus
correlation plot for the same data.
The dashed line is a linear fit
to the small-$C$ part of the data, with $\tf=0.65$.
The full line is the equilibrium FDT.}
\label{FDT1}
\end{center}
\end{figure}

The time-dependence of the dynamic functions $\chi_{\boldsymbol{k}}(t)$
and $[1-C_{\boldsymbol{k}}(t)]/T$ is displayed in
Fig.~\ref{FDT1}.
In that case, 200 realizations of the $\{\eps_j \}$
have been considered.
This figure shows that the equilibrium FDT is obeyed
at short times, and violated at larger times. From these
functions, a parametric plot is built, as shown in the inset of
Fig.~\ref{FDT1}.
It can be seen that the parametric plot can be described to an excellent
approximation by two straight lines, of slopes $-1/T$ (large
value of the correlation corresponding to short times) and
$-1/\tf$ (small value of the correlation
corresponding to large times).

This observation is thus in perfect agreement
with the behavior predicted theoretically~\cite{BBK}, and was
the main result in Ref.~\cite{BB}.

\subsection{Effective temperature in the plane
($\boldsymbol{\gamma,T}$)}

\begin{figure}
\begin{center}
\psfig{file=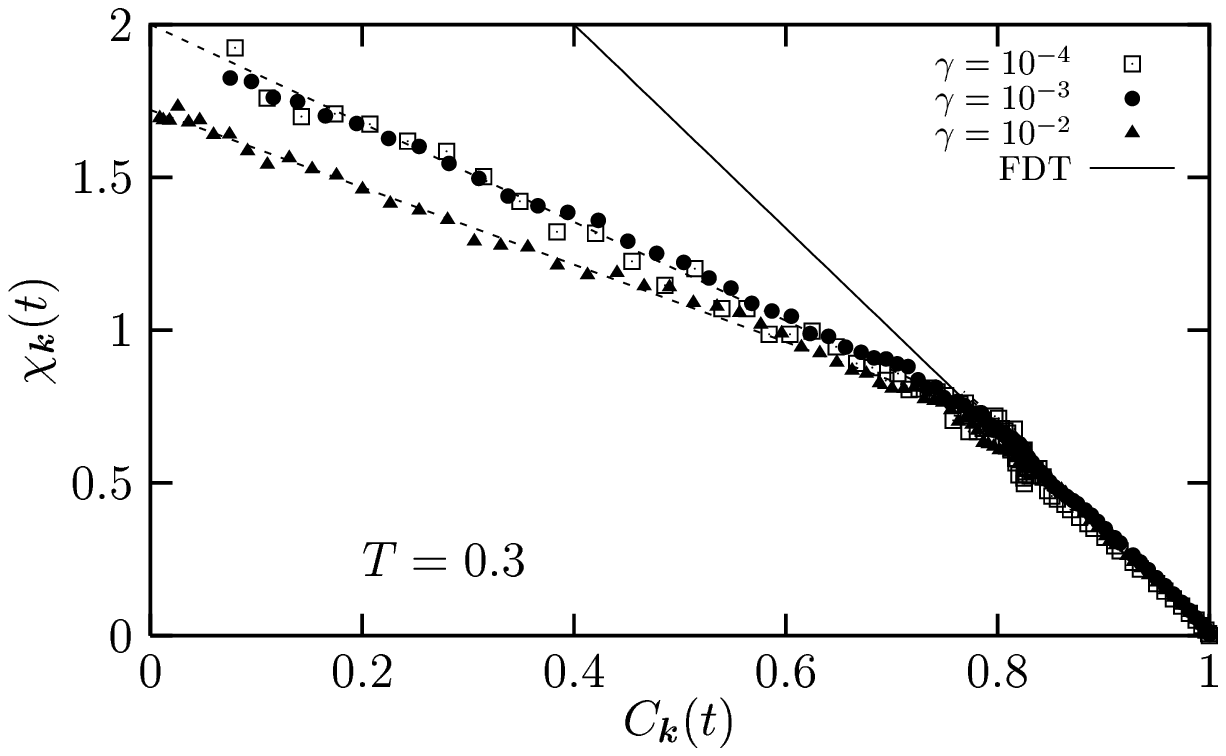,width=8.5cm,height=6cm} \\
\psfig{file=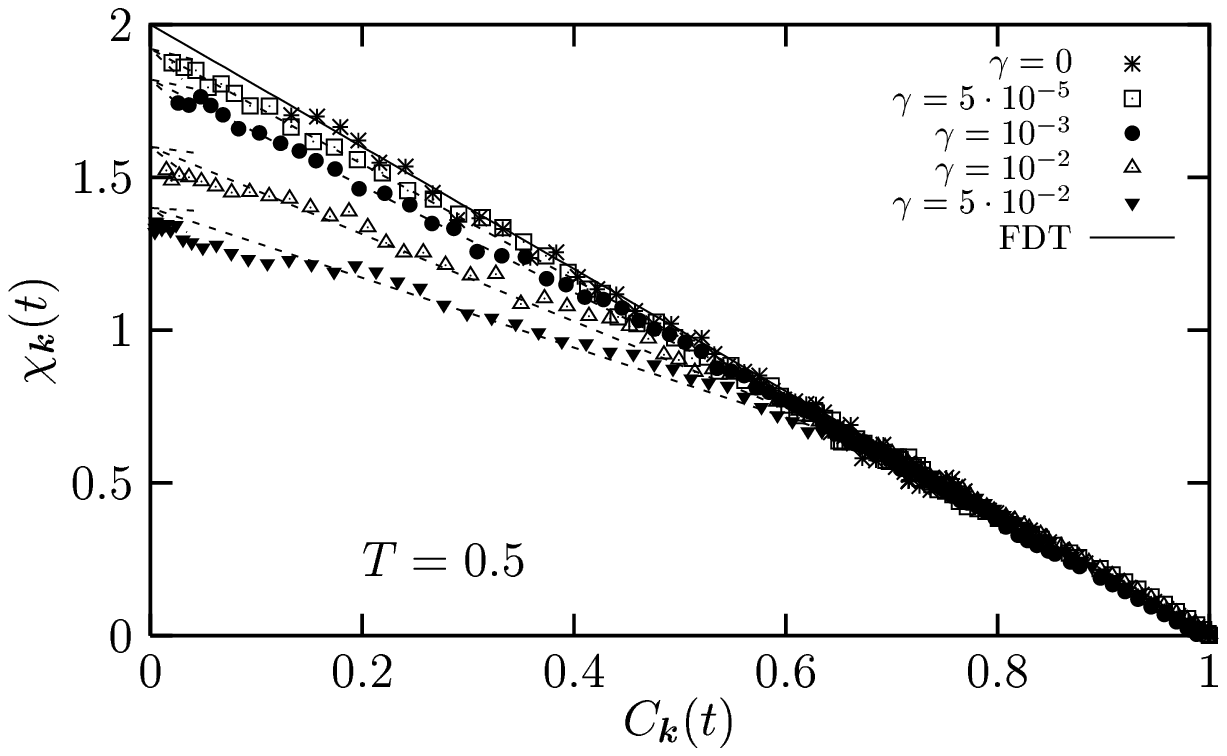,width=8.5cm,height=6cm}
\caption{Parametric plots for $T=0.3$ (top) and $T=0.5$ (bottom), and various
shear rates. In both figures, the full line is the FDT, and has a slope
$-1/T$.
The dashed lines are linear fits to the small-$C$ part of the data,
for $\gamma > 0$.
The wavevector is $\boldsymbol{k}=7.47\boldsymbol{e_z}$
for all the curves of this figure.}
\label{fig3}
\end{center}
\end{figure}

The effective temperature is defined as (minus) the inverse slope of the
parametric susceptibility-correlation plot for long times. This
effective temperature will, generally speaking, depend on how
strongly the system is externally driven. One may `reasonably'
expect that a stronger drive implies a higher effective
temperature. [See the discussion of the word `reasonably'
at the end of the paper.]
That this is indeed the case is demonstrated in
Fig.~\ref{fig3}, where parametric susceptibility-correlation
plots are displayed for several different shear rates.

Two different situations are considered in this figure. At high
temperatures, $T>T_c$, the effective temperature reduces to the
bath temperature when the shear rate is decreased. As predicted
theoretically~\cite{BBK}, we find here that the effective
temperature of the slow modes is equal to the bath temperature as
long as the system is in its Newtonian regime, $\eta \sim const$.
This is physically reasonable, since the linear regime of rheology
corresponds roughly to shear rates such that $\gamma \tr < 1$, for
which the dynamics is weakly affected~\cite{BBK,Onuki}.

In the `glassy' low temperature region, on the other hand, the effective
temperature also decreases with decreasing shear rate, but
saturates for the lower shear rates that can be
investigated in the simulation at a value different
from the bath temperature, as theoretically predicted~\cite{BBK}.
The existence of such a limiting value is physically
interpreted by the fact that at low temperature,
a Newtonian ($\eta \sim const$) or near-equilibrium ($\tf \sim T$)
regime defined as above
by the condition $\gamma \tr < 1$ does not exist, since
$\tr \sim \infty$.

The limiting value $\tf(\gamma \to 0)$
corresponds to the effective temperature that
was observed in the corresponding {\em aging} studies of the same
system at this temperature~\cite{barratkob}.
In Ref.~\cite{barratkob}, the limiting value $\tf \sim 0.62$ is
reported (however with much less accuracy than in the present paper),
while we find $\tf \sim 0.65$ at the same temperature $T=0.3$.
This is again expected from the theory~\cite{BBK}, and
confirms the interpretation of
$\tf$ as a signature of the geometry of phase space explored by
the system on the simulation time scale, independently of the
exact procedure which is followed~\cite{BBK,leticia3,Granular}.

\section{Observable independence of $\boldsymbol \tf$}
\label{obs}

As we already mentioned, the notion of an effective temperature
derived from the fluctuation-dissipation relation is much more
relevant if it can be shown not to depend on the observable under
consideration. That $\tf$ is observable independent is indeed one
of the crucial predictions of mean-field
approaches~\cite{latz,leticia3,doudou}. In this Section, we
explore this important issue by computing the
susceptibility-correlation parametric plots for several different
observables. Since the computation of response functions is
numerically very demanding, the method we used consisted in
computing very accurately the effective temperature from the
scattering function for $\boldsymbol{k}=7.47 \boldsymbol{e_z}$.
>From the data in Fig.~\ref{FDT1}, one has the estimation $\tf
\simeq 0.65$ for $T=0.3$ and $\gamma=10^{-3}$. We can then compute
for the same parameters ($T$, $\gamma$) the response associated
with other observables, and check whether these data are compatible
with this value of $\tf$. Since we have explained in some detail
the generic numerical procedure to measure $\tf$ in the previous
Section, less details will be given here. Instead, we carefully
discuss the possible experimental realizations of our
measurements. Note that in experiments,  a simultaneous
measurement of the response and  of the correlation is very
difficult, and has been  achieved only in very few
cases~\cite{grigera,ludovic,thesebellon,didier}. While usual
dielectric, magnetic or mechanical methods yield response
functions, scattering experiments  probe only correlation
functions~\cite{luca,lequeux}.

\subsection{Position correlations for different wavevectors}

\begin{figure}
\begin{center}
\psfig{file=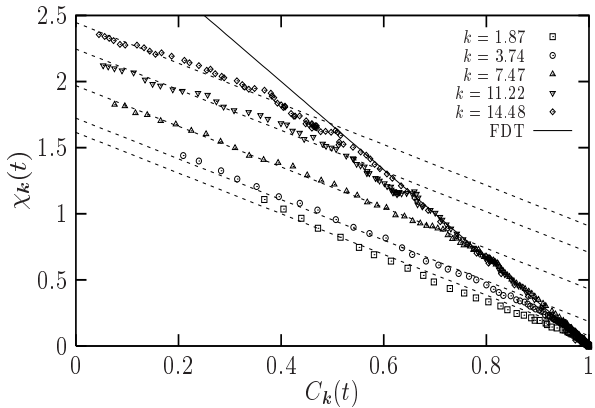,width=8.5cm,height=6cm}
\caption{Comparison of the parametric plot for $T=0.3$ and $\gamma=10^{-3}$,
and different wavevectors in the direction $z$.
The full line is the equilibrium FDT
of slope -1/0.3, the dashed lines have slope -1/0.65.}
\label{fig4}
\end{center}
\end{figure}

As a first check, we have investigated the wavevector dependence
of the parametric plot for the  self-part of the intermediate
scattering function. The results, displayed in Fig.~\ref{fig4},
are indeed perfectly consistent with a wavevector independent effective
temperature.
Parametric plots for different wavevectors differ
only through the value of the correlation function at which the
crossover from bath to effective temperature takes place. This
value marks the crossover between `short', equilibrated  and
`long', out of equilibrium  time scales. It also corresponds
to the plateau value $q_k$ in the correlation
functions displayed in Fig.~\ref{corr}.
Note that the wavevectors studied here cover a range of length
scales roughly between $0.5 \sigma_{AA}$ and $4 \sigma_{AA}$.
That these different length scales have the same off-equilibrium
behavior, as observed through the effective temperature,
is not a priori obvious.
It is in opposition, for instance, to the simple view of the system
rapidly equilibrating its small length scales, while its large
length scales need much more time to reach equilibrium, a
generic scheme which is put forward in systems such as
spin glasses~\cite{JP}.

\begin{figure}
\begin{center}
\psfig{file=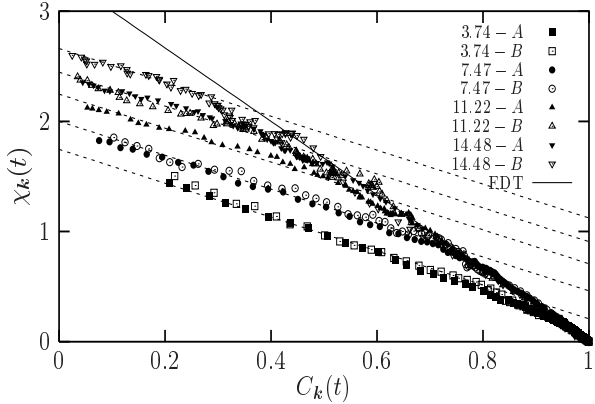,width=8.5cm,height=6cm}
\caption{Comparison
of the parametric plot for $T=0.3$ and $\gamma=10^{-3}$, for the two
types ($A$ and $B$) of Lennard-Jones particles. The data are
compatible with the same $\tf$ although the statistics for $B$
particles, the minority species, is slightly not as good as
for $A$ particles.
The full line is the equilibrium FDT
of slope -1/0.3, the dashed lines have slope -1/0.65.}
\label{fig5}
\end{center}
\end{figure}

The `chemical' dependence of the effective temperature can also
be investigated by computing the correlation function $C_{\boldsymbol k}(t)$
and the corresponding response, associated with $A$ and $B$ particles
of the Lennard-Jones binary mixture.
The results, displayed in Fig.~\ref{fig5}, again confirm that
the slope of the parametric plots does not depend on the particular
chosen observable, all the plots being well compatible
with the same $\tf=0.65$.

\begin{figure}
\begin{center}
\hspace*{-0.5cm}
\psfig{file=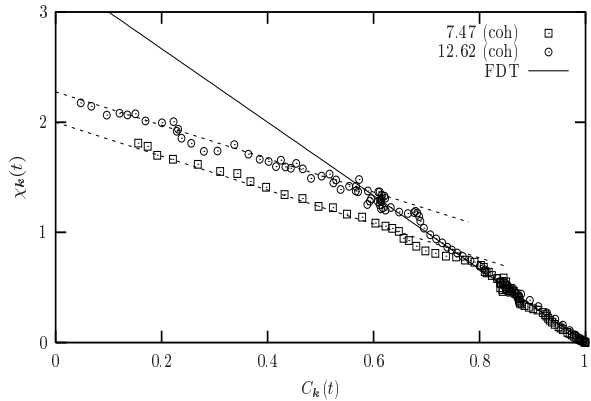,width=8.5cm,height=6cm}
\caption{Parametric plots from the coherent
density fluctuations for two
wavevectors $\boldsymbol{k}=7.47 \boldsymbol{e_z}$ and
$\boldsymbol{k}=12.62 \boldsymbol{e_z}$.
The full line is the equilibrium FDT
of slope -1/0.3, the dashed lines have slope -1/0.65.}
\label{fig8}
\end{center}
\end{figure}

Finally, the collective correlation and response functions can
also be computed ---albeit with a bit less accuracy than the
incoherent ones. This correspond to observables (\ref{obs1}) and
(\ref{obs2}) but taking $\eps_j \equiv 1$ for all $j$. The
resulting parametric plots are shown in Fig.~\ref{fig8}, for two
different wavevectors. Again, slopes are perfectly compatible with
equilibrium FDT for short times, and with $\tf = 0.65$ for longer
times.

It is interesting to consider how such
response-correlation plots could be obtained experimentally in
soft condensed matter systems. While the correlation functions are
reasonably easily obtainable through light scattering
experiments~\cite{luca,lequeux,pusey},
the same is not true of response functions. In order to obtain the
latter, one has to be able to manipulate the particles through
some externally applied potential, modulated
at the same wavevector as used in the light scattering experiment.
One suggestion here would be to
use some non-index matched tracer particles in an index matched
colloidal suspension. The tracer particles would then be sensitive
to the intensity of the local electric field. The same effect is
used, e.g., in optical tweezers.
An interference pattern would actually realize the
modulated external potential considered in this
Section. Reading of the response could then be obtained from the
scattering at the wavelength corresponding to this pattern.

\subsection{Einstein relation for the Lennard-Jones particles}

A quantity which is not directly related to the incoherent
correlation function is the mean square displacement of a tagged
particle.
The response function that is associated with this quantity is the
displacement induced by applying a small, constant external force to this
tagged particle. Both quantities are linked
by a fluctuation-dissipation theorem, called
the Einstein relation in that context.
To increase the statistics, we compute the mean square displacement
of all particles of the same type, for example
for the particle $A$, we measure
\begin{equation}
\Delta (t) = \frac{1}{2 N_A} \sum_{j=1}^{N_A} \left\langle \left[
z_j(t+t_0) - z_j(t_0) \right]^2 \right\rangle.
\label{msd}
\end{equation}
Again, we focus on the direction transverse to the flow,
for computational simplicity.
The conjugated response function is then
\begin{equation}
\chi_0(t) = \frac{1}{N_A F_0} \sum_{j=1}^{N_A} F_j z_j(t+t_0)
\label{repmsd}
\end{equation}
where $z_j(t)$ is the transverse position of particle $j$ at time
$t$. The force $F_j = \eps_j F_0$ is applied in
the direction $z$ between times $t_0$ and $t+t_0$.
The FDT reads then $\chi_0(t) = \Delta(t) / T$.

\begin{figure}
\begin{center}
\psfig{file=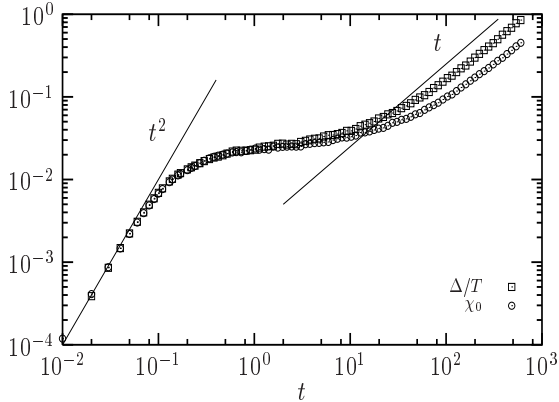,width=8.5cm,height=6cm}
\caption{Mean square displacement (\ref{msd}) normalized by the temperature,
and induced displacement (\ref{repmsd}) as functions of time.
At equilibrium, the FDT implies the equality of both quantities.
Full lines are the power law behavior of ballistic ($t^2$) and
diffusive ($t$) regimes.}
\label{fig6}
\end{center}
\end{figure}

\begin{figure}
\begin{center}
\psfig{file=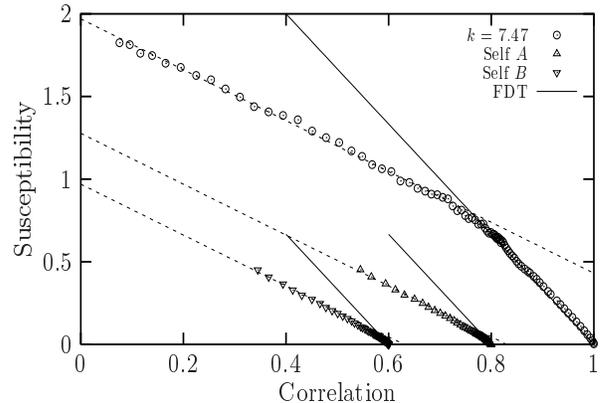,width=8.5cm,height=6cm}
\caption{Parametric plot for $T=0.3$ and $\gamma=10^{-3}$,
for the self-diffusion of the two types of Lennard-Jones particles.
We have represented the quantity $\chi_0$ versus $x-\Delta$, where
$x=0.6$ for $B$ particles, $x=0.8$ for $A$ particles,
for graphical convenience. These curves are
compared to the data obtained for the self part of the
intermediate scattering function at $\boldsymbol{k}=7.47\boldsymbol{e_z}$.
The full lines are the equilibrium FDT of slope -1/0.3,
the dashed lines have slope -1/0.65.}
\label{fig7}
\end{center}
\end{figure}

At short times, the mean square displacement is ballistic,
$\Delta(t) \sim t^2$. This ballistic regime is, as usual, followed
by a diffusive regime $\Delta(t) \sim t$.
A more extensive discussion of this quantity in the shear flow
may be found in Ref.~\cite{Onuki}.
The time behavior is similar for the induced displacement, as shown in
Fig.~\ref{fig6}. Hence, by construction, a linear relation is obtained
in the diffusive regime in the parametric plot of $\chi_0$ versus $\Delta$.
We confirm in Fig.~\ref{fig7} that, again, the effective temperature
defined in this way is compatible with the value $\tf=0.65$ already
obtained above. This is true for the two types of the Lennard-Jones
mixture, see Fig.~\ref{fig7}.

One may therefore define $\tf$ by simply using the relation
\begin{equation}
\tf = D \zeta
\end{equation}
between the diffusion constant $D$ and the friction coefficient $\zeta$
which are defined, as usually, as
\begin{equation}
\Delta(t) \sim  D t, \quad \chi_0(t) \sim \frac{1}{\zeta} t,
\end{equation}
for times which are in the diffusive regime.
This method was already used in Refs.~\cite{parisi,ruocco,sellitto} to
extract an effective temperature in aging systems,
and in Ref.~\cite{makse} in the context
of sheared granular materials.

Its interest, especially
in view of experimental realizations of these measurements, lies in the fact
that the full time dependence of the correlation and response
functions is not needed to extract the effective temperature.
Again, experiments could be considered if tracer particle
sensitive to an external force field (e.g. magnetic particles)
could be introduced into the system, and their mobility $\zeta$
measured together with their diffusion constant $D$.

\subsection{Stress fluctuations.}

A completely different observable, of relevance to all flow
situations, is the stress defined in Eq.~(\ref{stressmicro}).
In this Section, we study thus the case in which the two observables
$O(t)$ and $O'(t)$ are equal to the diagonal stress in the direction
perpendicular to the velocity and velocity gradient, $O(t) = O'(t)
= \sigma_{zz}(t)$.
To add a field conjugated to the normal stress, a
compression $\delta L_z$ of the simulation box is realized by
rescaling all particle coordinates at time $t=0$.
In the limit of small compression, $\delta L_z/ L_z \ll 1$,
the Hamiltonian $H$ of the system becomes indeed
\begin{equation}
H (\delta L_z) =  H(\delta L_z =0) + \frac{\delta L_z}{L_z}
V \sigma_{zz} + {\cal O} \left( \left(\frac{\delta L_z}{L_z}
\right)^2 \right),
\end{equation}
where $V=L_x L_y L_z$ is the volume of the system.
This proves that $\delta L_z L_x L_y$ is the field
thermodynamically conjugated to $\sigma_{zz}$ and allows
us to access another check of the observable independence of $\tf$.

\begin{figure}
\begin{center}
\psfig{file=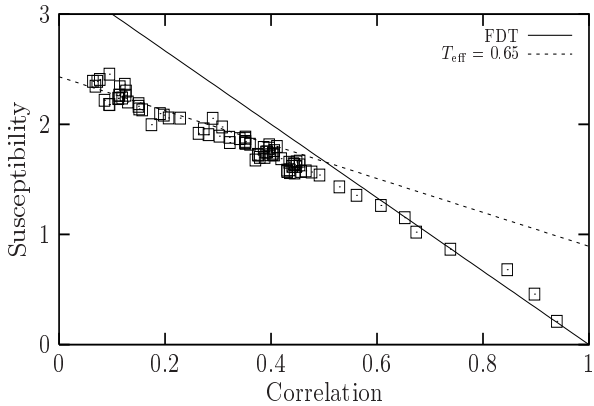,width=8.5cm,height=6cm}
\caption{Parametric plot for $T=0.3$ and $\gamma=10^{-3}$,
for the stress fluctuations.
The full line is the equilibrium FDT of slope -1/0.3,
the dashed line has slope -1/0.65.}
\label{stressparam}
\end{center}
\end{figure}

The resulting parametric plot is shown in Fig.~\ref{stressparam}.
Although the data are more noisy than in the density correlation case, the
FDT violations are very similar, and the value of $\tf$ is well
compatible with all the others.

Again, it may be possible to carry out the experiment
corresponding to this simulation. Experimentally, the off-diagonal
component of the stress (rather than a diagonal one) would be used
as the observable.  In fact, preliminary results in this
direction, using an extremely sensitive rheometer, have been
obtained on aging colloidal systems by Bellon and
Ciliberto~\cite{thesebellon}.
The superposition of a continuum flow in
order to obtain the fluctuations in a sheared system, however,
would represent a difficult experimental challenge.

\section{Further experimental proposals: Equipartition theorem
for the slow modes}
\label{propo}

As mentioned above, tracer particles can be used to probe the
dynamics of the fluid. In particular, the Einstein relation
between mobility and diffusion  leads easily to the definition of
an effective temperature, which appears to be identical to the one
obtained from various susceptibility-correlation functions.

A different use of tracer particles is considered in this Section,
in which we investigate the dynamics of tracer particles with a
mass large compared to that of the fluid particles. The idea is
the following. The definition of $\tf$ as a ratio between
correlation and response function has been shown to imply that, if
the fluid is used as a thermal bath to equilibrate a subsystem (or
`thermometer')  of typical time scale $t_s \sim \tr$, then the
Boltzmann weight for the subsystem  will be given by $\exp ( - E_s
/ \tf)$, with $E_s$ the energy of the subsystem~\cite{leticia3}.
The thermometer does not measures the microscopic temperature, but
the effective temperature associated with its characteristic time
scale. In that sense, $\tf$ can correctly be described as an
`effective temperature'.

Here, we propose a concrete realization of this situation in
which the thermometer is in fact a heavy particle, with a mass
$\mtr \gg m$.   Tuning the mass of the particle allows to control
its Einstein frequency $\omega_E$, defined by~\cite{hansen}
\begin{equation}
\omega_E  = \sqrt{\frac{\rho}{3 \mtr} \int \upd^3 \boldsymbol{r} \,
g(r) \nabla^2 V(r)},
\end{equation}
which essentially characterizes the frequency of vibration for the
particle in the cage formed by its neighbors. For particles of
type $A$ with $\mtr = 1$, $\omega_E \tau_0 \simeq 20$ at $T=0.5$.

In our simulations, we have considered   $10$ massive tracer
particles with masses in the range $\mtr \in [1,10^7]$. The tracer
particles are otherwise identical to $A$ particles,  so that the
equilibrium structure of the fluid is the same as with normal,
`light' particles. Since the oscillation frequency is inversely
proportional to the square root of the mass, the heaviest
particles will have an Einstein frequency typically thousand times
smaller than the light ones, which puts their oscillation period
in the range of typical relaxation times for the sheared
supercooled fluid. Note that the Einstein frequency is entirely
determined by static properties, and always determines the short
time behavior of the velocity autocorrelation of the particle.
Therefore it is a relevant quantity also for the description of
heavy particles, moving in a fast environment.

\begin{figure}
\begin{center}
\psfig{file=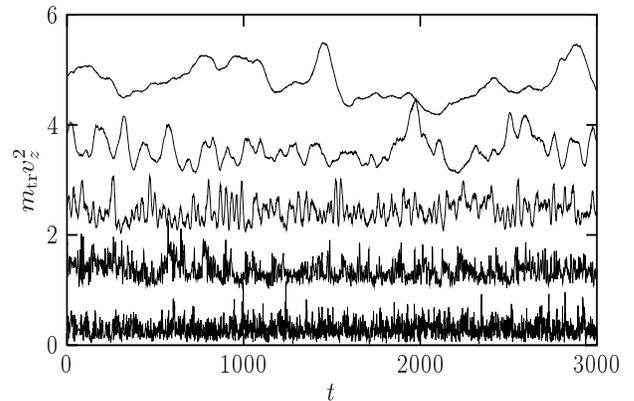,width=8.5cm,height=6cm}
\caption{Time dependence of the quantity $m_{\rm tr} v_z^2$
in particular trajectories of tracer particle of mass
$m_{\rm tr} = 10^2$, $10^3 \cdots 10^6$ (from bottom to top).
Curves have been vertically shifted for clarity.}
\label{tracer1}
\end{center}
\end{figure}

\begin{figure}
\begin{center}
\psfig{file=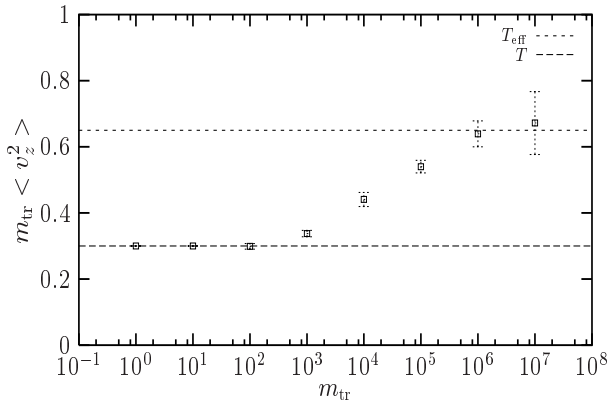,width=8.5cm,height=6cm}
\caption{Mass dependence of the mean kinetic energy in the $z$ direction for
$T=0.3$ and $\gamma=10^{-3}$.
Horizontal lines are $T=0.3$ and $\tf=0.65$.
Error bars are evaluated from tracer to tracer fluctuations.}
\label{tracer2}
\end{center}
\end{figure}

Fig.~\ref{tracer1} compares the typical trajectories for heavy
and light particles. The heavy particles are indeed seen to have a
much  lower oscillation frequency, and therefore their
dynamics can be seen as involving a low frequency filter coupled
to the fluctuations of the host fluid, which corresponds to the
slow, or `effective thermometer' considered in
Ref.~\cite{leticia3}. In order to get the temperature from this
thermometer, we compute the average  mean square velocity in the
direction transverse to the flow for these tracer particles. The
results are shown in Fig.~\ref{tracer2}.  From this figure, it
is clear that light particles have a mean squared velocity given
by the usual bath temperature ($T=0.3$ in this case), which
crosses over to the effective temperature $\tf=0.65$ for the heaviest
particles, $\mtr \sim 10^{6}-10^{7}$.
Intermediate masses, $\mtr \sim 10^{4}-10^{5}$ corresponds to times scales
located in the plateau region of the correlation which marks
the crossover between equilibrium and off-equilibrium behaviors.
This crossover is usually entirely hidden in the `breaking
points' near the value $C_{\boldsymbol{k}}  \sim q_k$ in
previous parametric plots.

\begin{figure}
\begin{center}
\psfig{file=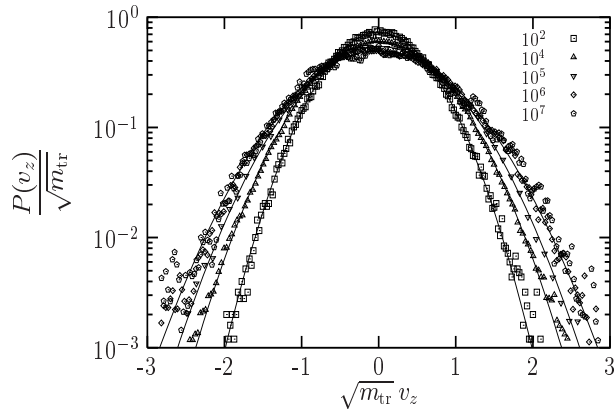,width=8.5cm,height=6cm}
\caption{Probability distribution function of the velocity $v_z$ for various
massive tracers, as indicating in the Figure.
The full lines are fits to the form (\ref{max}) demonstrating the
Gaussian shape for all the values of the mass
investigated. Values of $\tf$ are $\tf =0.3$
(equilibrium, $\mtr = 10^{2}$),
0.44, 0.54 (crossover, $\mtr = 10^{4}-10^{5}$) and
0.65 (asymptotic effective temperature, $\mtr = 10^{6}-10^{7}$).}
\label{maxwell}
\end{center}
\end{figure}

This implies that a {\it generalized equipartition theorem} holds
for the heaviest tracers, with the effective temperature
replacing the usual temperature, i.e.
\begin{equation}
\left\langle \frac{1}{2} m_{\rm tr} v_z^2 \right\rangle =
\frac{1}{2} \tf.
\end{equation}
Moreover, we show in Fig.~\ref{maxwell} that $P(v_z)$, the
probability distribution  of the velocity, is well approximated by
a Gaussian shape,
\begin{equation}
P(v_z) = \sqrt{\frac{m_{\rm tr}}{2\pi \tf}}
\exp \left[ - \frac{m_{\rm tr} v_z^2}{2 \tf}  \right].
\label{max}
\end{equation}
This in turn
implies that the usual Maxwellian shape of the velocity
distribution is recovered, again with $\tf$ replacing $T$.

The result, although expected on general grounds, is physically
quite surprising. Again, it could be tested against experiments
involving for instance colloidal particles, the tracers, in a
complex fluid, e.g. polymeric. The important factor here is not
the size of the tracer particle, but rather its mass that should
be much larger than that of the fluid constituants~\cite{note3}.
One could also investigate the rotational degrees of freedom of
non-spherical, `slow' tracer particles.

This result opens the way for new, simple determinations of
effective temperatures in out of equilibrium glassy materials.
Indeed, it has to be noted here that no `complex' dynamic
functions such as correlations or susceptibilities are needed.
This suggests that it would be extremely interesting to reproduce Perrin's
experiment~\cite{perrin} on barometric equilibrium in a sheared colloidal
suspension. Although we have not tested numerically this
situation, we expect indeed that the barometric equilibrium of
heavy particles inside a horizontally sheared fluid should also be
ruled by the effective temperature of the fluid, higher than the
room temperature.

\section{Discussion}
\label{discussion}

In this paper, we have investigated the correlation and response
functions for a supercooled fluid undergoing steady shear flow.
The ingredients essential to our study are the separation of time
scales existing in the fluid at rest between the slow
$\alpha$-relaxation and the microscopic times, and the energy input
at large scales provided by the shear flow. With these minimal
ingredients, it is possible to carry out a meaningful
comparison with theoretical results obtained for this
nonequilibrium dynamics at the mean-field, or schematic mode-coupling level.
We also believe that the
results should be of general relevance to many complex systems
undergoing shear flow, for which the same ingredients are present.

We have first studied the nonlinear rheology of the system, which
is indeed quite similar to what is observed in many complex
systems. The system exhibits shear-thinning for shear rates
exceeding the $\alpha$-relaxation time. The results can be
rescaled on a common curve as long as
temperature remains higher than the critical mode-coupling
temperature for this system.
Below $T_c$, the shear-thinning
behavior may alternatively be interpreted
in terms of a temperature-dependent shear-thinning exponent, or
a temperature-dependent yield stress.
The fact that $\sigma$ is nearly independant
of the shear rate at low temperatures and shear rates
makes the Lennard-Jones mixture
a good candidate to study possible mechanical instabilities
in Bingham fluids at a microscopic level.

In the shear flow, the system dynamics becomes stationary. This
allows an accurate determination of the time dependent correlation
functions, as opposed to the case of aging systems, for which
this determination is more difficult. We showed that several generic
features of correlation functions observed in supercooled systems
in the vicinity of $T_c$ are still present out of equilibrium,
as was predicted theoretically~\cite{BBK}.
Among those, the two-step ($\beta$ and
$\alpha$) relaxation,  the stretching of the $\alpha$-relaxation
and the factorization property which
characterizes the spatial dependence in the $\beta$-relaxation regime.

\begin{figure}
\begin{center}
\psfig{file=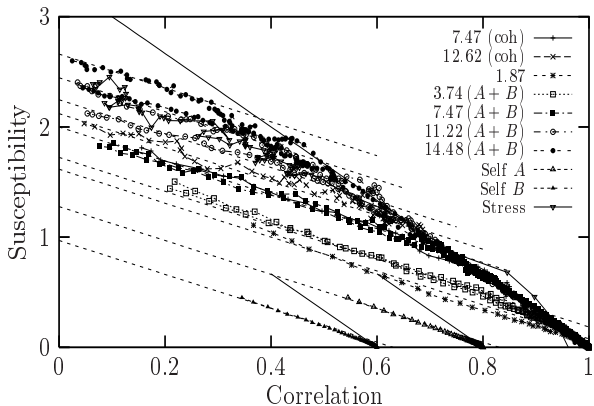,width=8.5cm,height=6cm}
\caption{Summary of our results for the effective temperature.
We show here the 14 different susceptibility-correlation parametric plots
described in the previous sections.
They are all consistent with the same effective temperature for the slow
modes. The full lines are the equilibrium FDT of slope -1/0.3,
the dashed line have slope -1/0.65.}
\label{spec}
\end{center}
\end{figure}

Finally, we have investigated in great detail the violations of
the fluctuation-dissipation theorem in this nonequilibrium system.
Our results, which  are summarized in a spectacular way in Fig.~\ref{spec},
are consistent with the expectation --based both on
theoretical arguments and on earlier results from aging studies--
that an effective temperature, independent of the observable under
consideration, can be defined for the slow modes of the fluid.
For several of the
observables we considered, we suggested experimental probes that
could be used to measure similar quantities in complex fluids.

We also presented a numerical realization of an `effective
thermometer' that probes this effective temperature, using heavy
tracer particles. The reading of the value of the effective
thermometer follows from the generalization of the equipartition
theorem to this nonequilibrium situation. This suggest a simple
and elegant method to access the effective temperature. Again, we
suggested experimental protocols that could realize these
measurements.

It has to be noted that
the notion of an effective (or fictive) temperature
is a long-standing issue in the context of glassy materials (see
e.g. Ref.~\cite{larson}). However, our results clearly show that the
definition Eq.~(\ref{definition}) from a nonequilibrium
fluctuation-dissipation has all the properties that has to be
captured by such a quantity~\cite{leticia3}.
(1) It is the temperature measured by a thermometer, and it
includes naturally the separation of time-scale of glassy systems,
as shown by our `tracer experiment'.
(2) It can be measured in a purely dynamic way, without making
reference to any equilibrated, or extrapolated, state.
(3) It is independent of the chosen observable, as we have clearly
demonstrated. None of these points is satisfied by previous definitions
of a fictive temperature~\cite{larson}.
Note that $\tf$ also captures the
essential phenomenological idea that when a system is sheared more
vigorously its effective temperature increases.
We thus emphasize here
the huge physical interest of this quantity, well beyond the
simple check of a sophisticated theoretical prediction.

All these results give strong support to
the theoretical scenario elaborated from schematic mode-coupling
theories to describe the rheology of soft glassy materials~\cite{BBK}.
This shows that although these theories give only a rather crude
description of a sheared fluid, they also capture its
essential features.
It is important to note that some of these features
are present in no other theory, such as the emphasis put on the
study of response functions, and the resulting FDT violations.

At the theoretical level, the next step
to be performed is of course to give a more `microscopic'
derivation of the nonequilibrium mode-coupling theory
of Ref.~\cite{BBK} for the study of glassy materials
under shear. In this respect, the results
obtained by Latz~\cite{latz} in its work on the aging regime
are a very important step.

However, we believe that it now is time to perform systematic
experiments to test quantitatively existing theories of the
nonequilibrium dynamics of glassy materials. In that sense, the
situation is more or less similar to the mid-eighties when
schematic equilibrium mode-coupling were already derived, but with
little experimental confirmation of its main features. This is why
we tried, as much as possible, to suggest experimental
counterparts to our  numerical measurements.  We hope that our
findings and suggestions in the present paper will motivate
further experimental work in the field.

\acknowledgements It is a pleasure to thank Jorge Kurchan and
Walter Kob for many useful discussions. This work was supported by
the P\^ole Scientifique de Mod\'elisation Num\'erique
at ENS-Lyon and the CDCSP at Universit\'e de Lyon.
PSMN and CDCSP are supported by the R\'egion Rh\^one-Alpes.


\begin{thebibliography}{199}
\bibitem{note1} Note that in previous works~\cite{kob,barratkob},
a different time unit, $\tau = (m_A\sigma_{AA}^2/48
\eps_{AA})^{1/2}$, is often used when dealing with Lennard-Jones particles.
\bibitem{note0} Strictly speaking, it is not impossible to observe
inhomogeneous flow under such conditions, see e.g.~\cite{roro} ; such
instabilities, however, seem to corresponds to inverse shear rates
of the order of microscopic oscillation times, higher than those
we are using.
\bibitem{note2}
As discussed in Ref.~\cite{roro}, potential problems
implying various thermostats  are expected only in fluid systems
where the shear rate is such that a particle moves by an
appreciable fraction of its diameter on a phonon period. In our
simulation, even the larger shear rate is ($\gamma=0.1$) is not in
this range.
\bibitem{note3} Obviously, it is not possible to achieve
experimentally the situation of large mass and small tracer size.
Our goal here was to realize in the simplest possible way a 'slow'
degree of freedom. As indicated in the discussion, we expect that
the important word here is  'slow'.  Actual tracers (brownian
particles), which are also 'large' particles, are expected to feel
the slow components of the fluid motion essentially in the same
way as the small, heavy tracers we are considering.
\end{thebibliography}
\end{document}